# Does Like Dissolves Like Rule Hold for Fullerene and Ionic Liquids?

Vitaly V. Chaban,[1,*] Cleiton Maciel,[2] and Eudes Eterno Fileti[3]

[1] MEMPHYS - Center for Biomembrane Physics, University of Southern Denmark, Odense M, 5230, Denmark

[2] Centro de Ciências Naturais e Humanas, Universidade Federal do ABC, 09210-270 Santo André, SP, Brazil

[3] Instituto de Ciência e Tecnologia, Universidade Federal de São Paulo, 12231-280, São José dos Campos, SP, Brazil

**Abstract**. Over 150 solvents have been probed to dissolve light fullerenes, but with a quite moderate success. We uncover unusual mutual polarizability of $C_{60}$ fullerene and selected room-temperature ionic liquids (RTILs), which can be applied in numerous applications, e.g. to significantly promote solubility/miscibility of highly hydrophobic $C_{60}$ molecule. We report electron density and molecular dynamics analysis supported by the state-of-the-art hybrid density functional theory and empirical simulations with a specifically refined potential. The analysis suggests a workability of the proposed scheme and opens a new direction to obtain well-dispersed fullerene containing systems. A range of common molecular solvents and novel ionic solvents are compared to 1-butyl-3-methylimidazolium tetrafluoroborate.

---

[*] Corresponding author. E-mail: vvchaban@gmail.com, chaban@sdu.dk (Vitaly V. Chaban).

TOC image

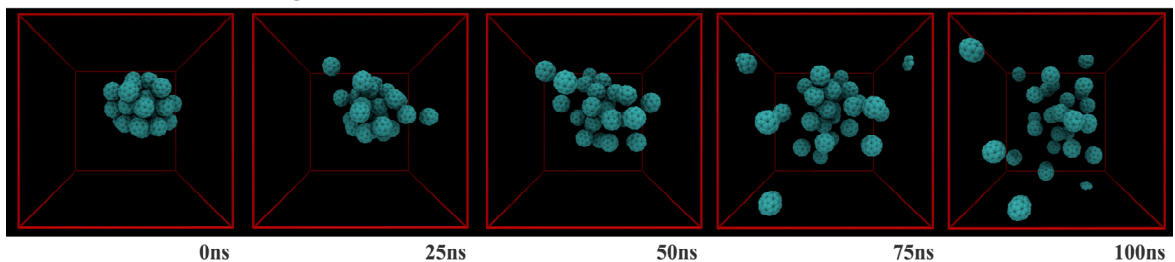

**Keywords**: fullerene, buckyball, ionic liquid, solvation, electronic structure, molecular dynamics.

## 1. Introduction

Efficient dissolution of fullerenes is an old quest started in 1985.[1] Fullerene solutions are important for numerous applications including, but not limited to, development of cosmetic products, robust drug delivery, extraction and separation of nanostructures from industrial soot, photovoltaics, investigation of chemical reactions involving fullerenes, etc.[2-10] Solutions and colloidal systems of fullerenes belong to one of the most studied systems in modern chemistry. A variety of solvents (over 150) and their mixtures were applied to obtain real solutions of $C_{60}$, $C_{70}$, and some higher fullerenes.[4, 11, 12] The available data are, however, rather pessimistic. In the majority of polar solvents, including water, solutions of fullerenes are either extremely diluted or colloidal ones.[4] Relatively concentrated solutions are achieved in some solvents of low polarity, such as aromatic hydrocarbons and their haloderivatives. For instance, solubility of $C_{60}$ in benzene equals to 1.50 g l$^{-1}$,[13] 2.40 g l$^{-1}$ in toluene,[14] 2.60 g l$^{-1}$ in ethylbenzene,[13] all measurements being done at 298 K. Remarkably, methylation of the solvent molecule significantly increases its affinity to $C_{60}$, although chemical background of this empirical finding is unclear. 1,2,3-trimethylbenzene allows for solubility increase by more than three times, compared to benzene, up to 4.70 g l$^{-1}$.[15] The same authors reported 17.90 g l$^{-1}$ for 1,2,4-trimethylbenzene and 5.80 g l$^{-1}$ for 1,2,3,4-tetramethylbenzene,[15] although the first quantity looks doubtful for us. Halogenation of low molecular aromatic compounds provides even better results in many cases: 6.35 in chlorobenzene, 13.80 in 1,2-dibromobenzene, 9.62 in 1,2,4-trichlorobenzene, etc.[12] Grafting other functional groups does not lead to comparable effects. Benzonitrile, nitrobenzene, phenyl isocyanate, n-propylbenzene, *n*-butylbenzene and a few other solvents, carefully summarized by Semenov et al.,[11, 12] are notably unsuccessful for pristine light fullerenes.

The highest solubility of $C_{60}$ that we are aware of was achieved by Talukdar and coworkers,[14] who used piperidine (53.28 g l$^{-1}$) and pyrrolidine (47.52 g l$^{-1}$). It was hypothesized that these solvents exhibited specific donor-acceptor interactions with $C_{60}$, $C_{70}$, $C_{84}$, $C_{100}$, that can, under certain conditions, give rise to formation of new chemical compounds. Piperidine and pyrrolidine are sometimes called "reactive" solvents with respect to $C_{60}$. Note, that published references on solubility contain lots of contradicting results. This fact can be correlated with specific methods of measurements, their internal uncertainties, and preparation of fullerene containing solutions.[4, 12, 16-18] In all polar and many low-polarity solvents, fullerenes tend to form colloidal systems, being another source of the scattered, ambiguous results.

Apart from direct experiments outlined above, numerous computer simulations have been conducted on fullerenes in solutions.[19-22] Monticelli and coworkers published a few important developments in atomistic and coarse-grained force fields,[19, 23, 24] one of us reported free energies and structure patterns of $C_{60}$ in various solvents,[25-27] a few high-impact works were devoted to the partitioning of $C_{60}$ and its derivatives at the water/bilayer interface.[21, 28] Density functional theory (DFT) was applied to probe applicability of fullerenes in photovoltaics and obtaining fundamental understanding of their electronic structures.[10, 29-32] In the meantime, we are not aware of any investigations of the fullerene dissolution process in real time and length scales.

To summarize, even if "reactive" solvents are counted, pristine fullerenes cannot currently be solvated in the same sense as conventional, smaller molecules can be. Unlike oxygens in ozone, carbons in fullerenes are chemically equivalent and, therefore, no permanent electric moment exists. Short-range dispersive attraction in case of carbon is inferior to that of more electronegative elements, but is strong enough to keep fullerenes in a solid state at ambient conditions. Because of the supramolecular size of fullerene, its introduction into solvent makes entropic part of solvation free energy, $TdS$, significantly negative, whereas enthalpic constituent is never high enough. The time has come to look at dispersion of fullerenes from a new molecular perspective.

In the present work, we introduce an alternative concept of solvating fullerenes using room-temperature ionic liquids (RTILs). Our concept utilizes genuine high electronic polarizability of fullerenes, as compared to other carbonaceous compounds, and peculiarities of delocalized electron density in organic cations of RTILs. The original evidence of our method efficiency was obtained from functional density functional theory coupled with a highly accurate hybrid exchange-correlation functional, and large-scale atomistic-precision molecular dynamics (MD) simulations. It follows from our modeling and simulations that a classical "Like Dissolves Like" rule does not hold for the case of fullerene and 1-butyl-3-methylimidazolium tetrafluoroborate.

**2. Insights into non-bonded fullerene-ionic liquid interactions**

In their condensed state, most ionic liquids[33-35] are very polarizable and there is an electron density transfer between anion and cation.[36-38] We selected one of such RTILs, 1-butyl-3-methylimidazolium tetrafluoroborate, [BMIM][BF$_4$], to demonstrate our new concept. In turn, fullerenes are also known for high dipole polarizability, 80 Å$^3$,[39] because of unusual electronic structure and icosahedral symmetry. In order to measure the extent of nonadditivity of C$_{60}$-RTIL non-bonded interactions, we employed hybrid density functional theory methodology. We describe an electron density using a high-quality hybrid exchange-correlation functional, omega B97X-D.[40] Within this functional, the exchange energy is combined with the exact energy from Hartree-Fock theory. Furthermore, empirical atom-atom dispersion correction is included, being of vital importance in case of fullerenes. Hybrid functionals have been shown to bring significantly improved accuracy in the resulting electronic structure of the simulated quantum mechanical systems, and therefore, polarization effects. While many developments, in particular based on local density approximation, tend to overestimate electron transfer (ET) between nuclei, carefully "trained" hybrid functionals are free of this defect. According to Chai and Head-Gordon, the functional simultaneously yields satisfactory accuracy for thermochemistry, kinetics, and non-covalent interactions. Based on numerous tests, omega B97X-D is superior for non-bonded interactions over previous hybrid functionals.[40] The 6-31G(d) basis set, containing vacant *d*-orbitals, was used in all computations, since this basis set is continuously applied to describe a great variety of carbonaceous compounds. Whereas 6-31G(d) is less comprehensive than Dunning's correlation-consistent basis sets (especially triple, quadruple, and higher order implementations[41]), it provides reasonable accuracy in case of ground-state systems. Note that implementation of 6-31G(d) instead of aug-cc-PVTZ[41] decreases computational cost of the same hybrid DFT method by at least one order of magnitude.

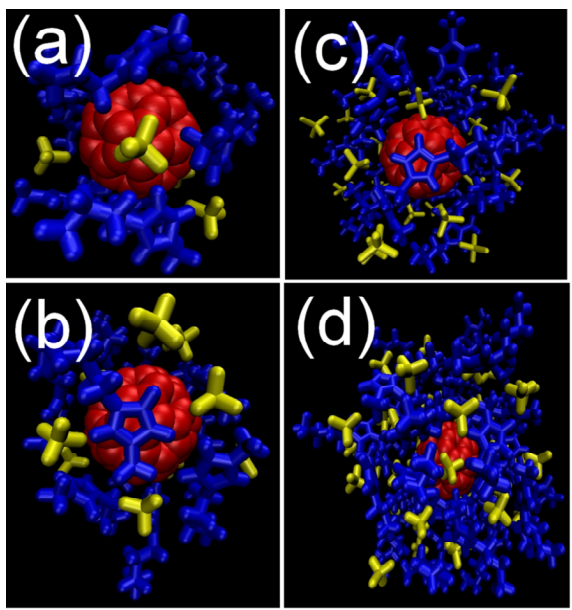

Figure 1. The first solvation shells of fullerene. All ions of the shell are within 0.8 (a), 1.0 (b), 1.2 (c), and 1.4 nm (d) from $C_{60}$.

The system containing one $C_{60}$ and 500 [BMIM][BF$_4$] ion pairs was originally simulated using empirical scaled-charge potential for [BMIM][BF$_4$], introduced by one of us earlier.[36] The $C_{60}$ fullerene was represented by suitable AMBER-type Lennard-Jones (12,6), LJ, carbon parameters.[42] The set of empirical parameters for $C_{60}$ and [BMIM][BF$_4$] is listed in Ref.[43] After equilibrium had been reached at 300 K, we extracted coordinates of $C_{60}$ with its solvation shells. Solvation shells within the radius of 0.8 (5 ion pairs), 1.0 (10 ion pairs), 1.2 (24 ion pairs), and 1.4 nm (34 ion pairs), Figure 1, were generated. $C_{60}$ in contact with a single ion pair was used for comparison. Binding energy, corresponding to non-bonded interactions in each shell, electron localization, and full orbital structure were calculated using omega B97X-D/6-31G(d). Basis set superposition error (ca. 20% of average binding energy) was removed using the counterpoise technique. Figure 2 suggests that certain electron transfer indeed occurs between ions and $C_{60}$. Brought into direct non-bonded contact with RTIL, fullerene lost a fraction of electron density, resulting in the +0.35e charge. The depicted charges were obtained using real-space integration of electron density within certain radius from the nucleus, based on the classical Hirshfeld technique. The total electron charge on $C_{60}$ stabilizes after number of ions achieves the size of the fullerene first solvation shell. Additional increase of the shell size does not lead to any alteration of the electron transfer phenomenon. The binding energies (Figure 3) between $C_{60}$ and RTIL suggested by original potential (essentially based on the Lorenz-Berthelot assumption) appear underestimated by 15-20%. We correlate this underestimation with the electron transfer discussed above (Figure 2). The transfer is promoted via disturbance of the fullerene highest energy electrons by an electric field of neighboring ions. However, a complete ionization of $C_{60}$ does not take place in these systems. Based on the same modeling scheme, the cationization of $C_{60}$ in vacuum costs the system ca. 400 kJ/mol, whereas anionization costs ca. 500 kJ/mol. Since the observed non-additive interaction is caused by induced dipole, it should be described by a specific energetic well depth parameter. Such a parameter can be epsilon in LJ potential, in contrast to adjustment of permanent Coulombic charges on the $C_{60}$ and ions. Using an iterative procedure, we altered epsilons for pairwise $C_{60}$-$BMIM^+$ and $C_{60}$-$BF_4^-$ interactions. An optimal set of parameters to approximate non-bonded energetics, predicted by DFT, was generated (Figure 3). Note, that well depth parameters were changed uniformly, irrespective of nature of each chemical element. It is an obvious approximation, justified by the fact that all involved atoms are of similar size and therefore exhibit similar polarizabilities. Pairwise epsilons were changed only for polar moieties of RTIL. The workability of this approximation to accurately describe (Figure 3) binding in case of each of the five shells is a key evidence in favor of our methodological solution. After the refinement of the force field, the MD system was additionally equilibrated, new solvation shells were produced, and DFT calculations were repeated. However, the resulting binding energies were nearly the same as in the initial case. The solvation shell structures and ion orientations near the fullerene surface remained intact. According to snapshots in Figure 1, $C_{60}$ is preferentially coordinated by cations. Interestingly, cations coordinate $C_{60}$ by their polar moiety, imidazole ring, and not by the hydrophobic -CH$_2$-CH$_2$-CH$_2$-CH$_3$ chain. The qualitative observations were the same, irrespective of using original FF or refined FF. This orientation is more preferable for a periodic system due to relatively strong dispersion-type attraction between nitrogen atoms of imidazole ring and carbon atoms of fullerene. Although RTIL imposes certain polarization on $C_{60}$, no *chemical bonding* occurs between these particles. In case of the first solvation shell (10 ions pairs) the binding energy is 30 kJ/mol at 300 K per one [BMIM][BF$_4$] ion pair. Being normalized per number of atoms in the polar part of cation and anion, it is weaker than conventional hydrogen bonding. A few highest energy π-orbitals of $C_{60}$ are responsible for the observed electron transfer. Figures 2 and 3 predict that 0.1 to 0.35e are transferred from fullerene to anion (i.e. to more electronegative fluorine atoms). Noteworthy, the second solvation shell (see values corresponding to 24 and 34 ion pairs) does not contribute to electron transfer. In turn, the number of ions in the first solvation shell plays a major role (see values corresponding to one, five, and ten ions pairs). It would be instructive in future investigations to gain insights about charge transfer dynamics and extend studies to another temperature range, since electronic polarization is known to depend on temperature. Remarkably, solvation leads to energy increase by 0.1-0.3 eV (depending on the shell size) of the highest occupied and the lowest unoccupied orbitals, localized on fullerene. Such significant shift normally indicates a strong coupling between the solute and the solvent.

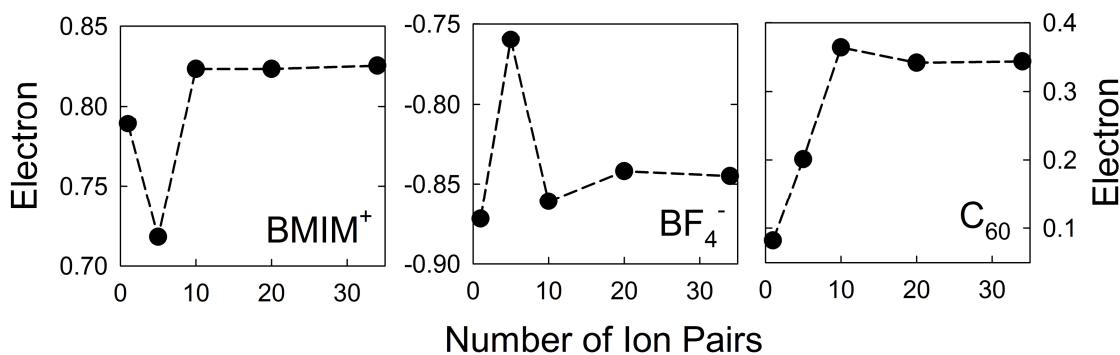

Figure 2. Electron charge on cation ($BMIM^+$), anion ($BF_4^-$), and fullerene ($C_{60}$) as a function of ion pairs number.

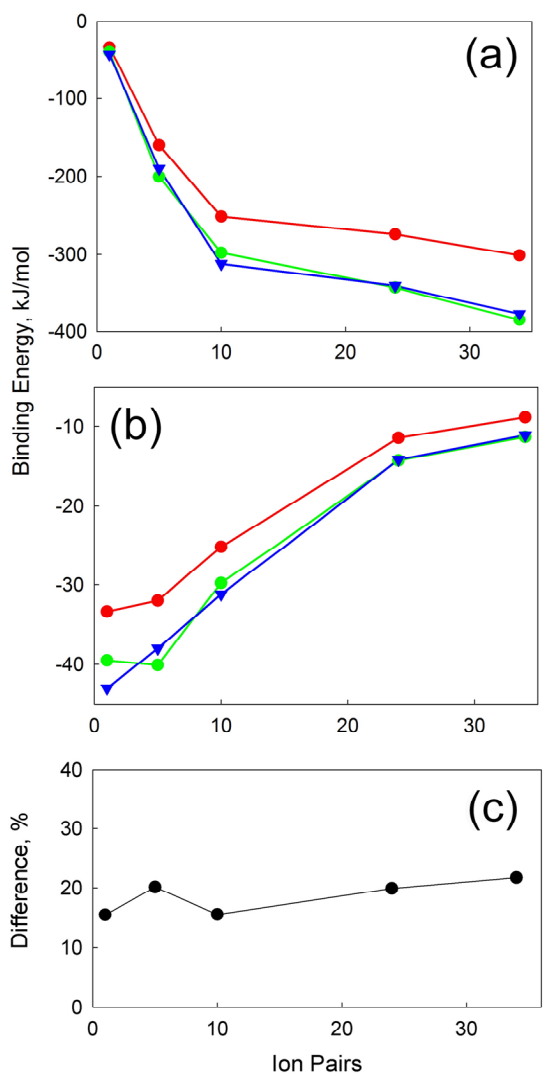

Figure 3. Binding energy between $C_{60}$ and [BMIM][BF$_4$] as a function of ion pairs number: (a) normalized per mole of fullerenes; (b) normalized per mole of [BMIM][BF$_4$]; (c) percentage difference between original and refined models. Original model is red circles, DFT energy is green circles, refined model is blue triangles.

### 3. Insights from large-scale molecular dynamics

A solid species of the $C_{60}$ fullerene was immersed into liquid 1-butyl-3-methylimidazolium tetrafluoroborate (1500 ion pairs), Figure 4. Non-equilibrium molecular dynamics was simulated at 300, 310, 320, 350, and 400 K under ambient pressure, until the systems came to thermodynamic equilibrium. Equilibrium properties were derived using 30 ns long trajectories with a time-step of 0.002 ps and coordinates saving frequency of 20,000 frames per nanosecond. Velocity rescaling thermostat[44] and Parrinello-Rahman barostat[45] with coupling constants of 1.0 ps and 4.0 ps were turned on to represent NPT ensembles of these systems. Covalent bonds of heavy elements with hydrogen atom were constrained using the LINCS algorithm,[46] which allows for integration time-step increase without influencing computational stability. The force field was refined as described above. Note, that our model provides average $C_{60}$-RTIL and RTIL-RTIL energetics ideally concordant with average energetics from hybrid density functional theory method. The Lennard–Jones interactions were gradually shifted to zero between 1.2 and 1.3 nm. The real-space Coulomb interactions were truncated at 1.4 nm. Their long-range parts were taken into account via the particle-mesh Ewald method.[47] All large-scale simulations were carried out using the GROMACS 4 software package.[48-51]

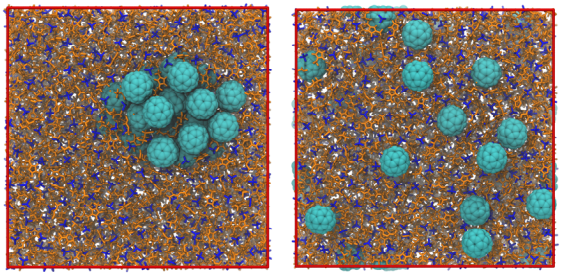

Figure 4. Representative configuration snapshots of the solid (left) and dissolved (right) state of $C_{60}$.

Fullerene $C_{60}$ is nearly insoluble at 300 K. Although the well-ordered periodic structure of the solid phase changes to cluster-like (Figure 4), no fullerene is present in dissolved state (Figure 5) for a significant time during 300 ns of the spontaneous evolution of MD system. This result is in excellent agreement with recent experimental observations,[52] where a few popular RTILs were probed for the $C_{70}$ solvation at room temperature. None of the RTILs, such as 1-butyl-2,3-dimethylimidazolium bis(trifluoromethylsulfonyl)imide, 1-methyl-3-octylimidazolium tetrafluoroborate, 1-methyl-3-octylimidazolium hexafluorophosphate, 1-methyl-3-octylimidazolium bis(trifluoromethylsulfonyl)imide, 1-decyl-3-methylimidazolium tetrafluoroborate, methyltrioctylammonium bis(trifluoromethylsulfonyl)imide, trihexyltetradecylphosphonium chloride, etc, appeared an efficient solvent for fullerene, according to fluorescent measurements reported by Martins et al.[53] However, some of the above RTILs demonstrated better solvation of $C_{70}$ as compared to [BMIM][BF$_4$], from 0.02 up to 0.06 g l$^{-1}$. Unfortunately, all cited values are nearly negligible.

Despite failure at ambient conditions, [BMIM][BF$_4$] becomes outstandingly successful upon just 20 K temperature increase, 5 g l$^{-1}$ (310 K), 49 g l$^{-1}$ (320 K), > 66 g l$^{-1}$ (333 K). Since enthalpic factor does not change significantly between 300 and 320 K,[9] the dissolution is boosted due to entropic factor increase. We were unable to identify *maximum* $C_{60}$ solubility at higher temperatures, since the originally scheduled systems contained only 30 fullerene molecules corresponding to a maximum solubility of 66 g l$^{-1}$. This limit was achieved and exceeded at 333 K. Compare 66 g l$^{-1}$ to the solubility potential of "reactive" solvents.[14] No reaction between $C_{60}$ and RTIL occurs in our case.

The above findings look very encouraging and motivate to take a closer look at the representatives of other families of RTILs in order to foster solubility towards ever higher values. Mutual polarization of [BMIM][BF$_4$] and $C_{60}$ plays a key role in successful dissolution. Ionic nature of the [BMIM][BF$_4$] liquid and delocalized charge on the cation are important prerequisites for their electronic polarizability. The elaborated approach is possible only with RTILs, but impossible with solutions of ions in *polar* molecular liquids. In the latter case, ions are strongly solvated by the solvent molecules and cannot approach the fullerene. The fullerene is surrounded predominantly by neutral particles, which, in most cases, exhibit insignificant (e.g. alcohols) to negligible (e.g. nitriles) polarization potential.

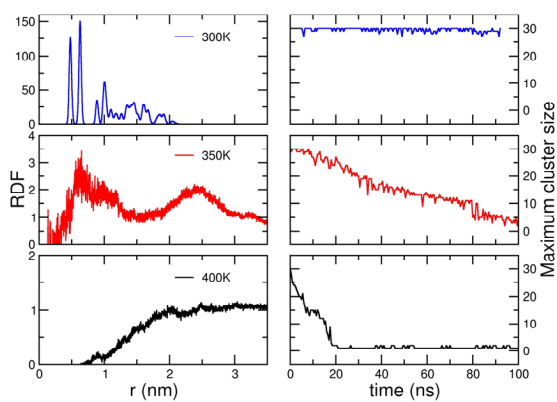

Figure 5. (left) Radial distribution function of the center of mass of fullerenes; (right) evolution of the largest cluster size versus simulation time.

Figure 5 depicts radial distribution functions between fullerene center-of-masses. Undissolved cluster at 300 K (Figure 4) is evidenced by sharp, well-ordered peaks. The arrangement of peaks in not ideal though, as it is observed at lower temperatures, below 260 K (not shown). Therefore, [BMIM][BF$_4$] exhibits a strong potential for breaking the $C_{60}$-$C_{60}$ weak bonds even at ambient conditions, and insignificant changes of entropic or enthalpic factors can alter the solvation behavior drastically. It happens at higher temperatures (e.g. 320-333 K), resulting in the unexpectedly high value of minimum 66 g l$^{-1}$. Larger clusters of $C_{60}$ must be simulated at T > 333 K to tabulate solubility. On a related note, Li and co-workers have recently conducted Raman and IR measurements, as well as DFT calculations, to investigate the dispersion of single-walled carbon nanotubes in imidazolium-based ionic liquids at room temperature.[53] On the contrary to their expectations, all methods unequivocally suggested no strong interaction, such as π- π, between carbon nanotubes and imidazolium cations. The only recorded interactions were weak van der Waals attraction, indicating that RTILs are promising solvents to disperse nanotubes without influencing their electronic structure and properties.

The distance of 0.8 nm defines the first coordination sphere of the cluster at 300 K (Figure 5, left). The neighboring fullerene molecules are considered aggregated, when the distance between their centers does not exceed 0.8 nm. With this information in mind, it was possible to analyze the time evolution of the dissolution process in terms of the size of the largest cluster (Figure 5, right). The dissolution process takes ca. 20 ns at 400 K and ca. 100 ns at 350 K. The dissolution time/speed depends on the volume of solid phase and on interface surface. With the above data from MD simulations, the speed of $C_{60}$ solvation can be assessed for various aggregate sizes and various external conditions. Statistics of the cluster sizes (Figure 6), accumulated during an equilibrium stage of simulation, clearly illustrates that dissolution takes place. This process is thermodynamically favorable. Interestingly, a few $C_{60}$ dimers and trimers are recorded in [BMIM][BF$_4$], although their fractions do not exceed 10%. Such dimers and trimers are wide-spread in the aqueous and nonaqueous electrolyte solutions.

It should be underlined that our observations break one of the most solid rules in chemistry, the "Like Dissolves Like" rule. This statement indicates that a solute will dissolve best in a solvent that has a similar chemical structure to itself. The solvation capacity of a solvent depends primarily on its polarity, i.e. dipole moment. For instance, a hydrophilic solute, such as sodium sulfate, is soluble in polar water, less soluble in fairly polar ethanol, and insoluble in all non-polar solvents, such as benzene. In contrast, a hydrophobic solute, such as naphthalene, is insoluble in water, fairly soluble in ethanol, and soluble in non-polar benzene. The fullerenes are well-known as highly hydrophobic solutes, whereas 1-butyl-3-methylimidazolium tetrafluoroborate is a polar ionic compound. According to the rule, they should not form a single phase, but in practice they form quite a concentrated solution, thanks to the electron transfer phenomenon. One can rephrase that [BMIM][BF$_4$] lowers hydrophobicity of $C_{60}$, after they mix.

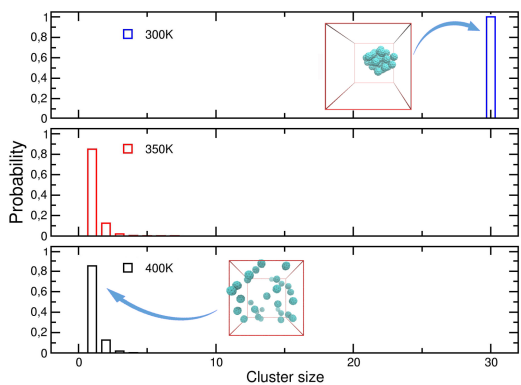

Figure 6. Cluster size distributions of $C_{60}$ in [BMIM][BF$_4$] at 300, 350, and 400 K. Whereas all fullerenes belong to the same cluster at 300 K, a variety of aggregates (monomers, dimers, trimers) are observed.

### 4. Ionic liquids versus molecular liquids

In order to generalize our findings, we performed an additional investigation of the electron transfer effect for a set of ionic liquids, representing four families. The solvation shells, consisting of N-butylpyridinium chloride ([BPY][Cl]), ethylammonium bis(trifluoromethane)sulfonamide ([Et$_4$N][TFSI]), ethylphosphonium tetrafluoroborate ([Et$_4$P][BF$_4$]), and 1-methyl-3-ethylimidazolium hexafluorophosphate ([EMIM][PF$_6$]), were considered using the methodology introduced above. The obtained results were compared to a range of popular molecular solvents, such as ethanol, acetonitrile, dimethyl sulfoxide (DMSO), cyclohexane, 1,2,4-trimethylbenzene, chloroform, whose modest to average performance for $C_{60}/C_{70}$ dispersion had been already documented.[12] The electron transfer in each system is summarized in Table 1. According to hybrid DFT, $C_{60}$ exists as a partially charged particle in nearly all solutions. The absolute value of total charge on $C_{60}$ varies from 0.01 e (acetonitrile and [Et$_4$N][BF$_4$]) to 0.47 e ([BPY][Cl]). We did not find a clear correspondence between polarity (dipole moment) of the solvent and total charge on the $C_{60}$. For instance, acetonitrile possesses a significant dipole moment (3.9 D), but it imposes a negligible total charge on $C_{60}$. In turn, dipole moment of [BPY][Cl] ion pair is large, 6.12 D (computed value). Cyclohexane molecule is non-polar, does not contain either strongly electronegative elements or aromatic moieties, nevertheless ET equals to 0.16. We suppose that ET can be numerically correlated with HOMO and LUMO energy levels in combination with localization of these orbitals. The hypothesis needs more comprehensive elaboration with a synergetic application of ab initio modeling and spectroscopic experiments.

Table 1. Electron transfer from $C_{60}$ fullerene to its solvation shell formed by ionic and molecular solvents, computed using hybrid density functional theory at the omega B97XD/6-31G(d) level.

| Ionic Liquid | Electron transfer, e | Molecular Liquid | Electron transfer, e |
| --- | --- | --- | --- |
| [EMIM][PF$_6$] | 0.18 | Ethanol | 0.06 |
| [Et$_4$N][TFSI] | 0.09 | Acetonitrile | 0.01 |
| [Et$_4$P][BF$_4$] | 0.01 | DMSO | 0.07 |
| [BPY][Cl] | **0.47** | Chloroform | 0.23 |
| [BMIM][BF$_4$] | **0.35** | Cyclohexane | 0.16 |
| | | 1,2,4-trimethylbenzene | 0.08 |
| | | water | 0.27 |

Ammonium and phosphonium RTILs do not polarize $C_{60}$, irrespective of the anion. The most intuitive explanation is that positive charge on the cations is deeply buried and is chemically inactive. Interestingly, anions do not impose any effect on the fullerene in these two cases. However, a transfer of 0.35 e was recorded when tetrafluoroborate is coupled with an imidazolium cation.

Among molecular solvents, water and chloroform exhibit the largest ETs, 0.27 and 0.23, respectively. Both liquids are polar solvents, μ ($H_2O$)=1.85 D, μ ($CHCl_3$)=1.04 D, with a significant degree of dipole-dipole interactions determining their molecular structures in condensed phase. The van der Waals interactions are not decisive in chloroform and marginal is water. Despite the derived ET, these solvents cannot be efficient for fullerenes. This conclusion is well supported by direct experiments on solubilization.[12] Thus far, electron transfer effect was not included in computer simulations of $C_{60}$ dispersibility in the liquid phases, since it was assumed insignificant. Accounting for ET in future must help to achieve better description of structure and transport properties of the mentioned systems.

1,2,4-trimethylbenzene was reported as a relatively successful solvent before (even better than benzene),[12] but the computed ET is a few times smaller than that of pyridinium and imidazolium RTILs (Table 1). The good solvation ability of 1,2,4-trimethylbenzene probably comes mainly from entropic contribution to the solvation free energy and not from the electronic polarization.

Solvation is observed everywhere in nature. Most of chemical reactions occur in solutions. This is an extremely complicated process involving a number of independent and semi-independent factors, such as strength of dipole-dipole interactions, π-π interactions, weak dispersion interactions, hydrogen bonding, molecular and ionic shapes, conformational flexibility, potential energy surfaces, ability of solute to dissociate, mass and electron densities, temperature, external pressure, etc. These factors are classically divided into entropic and enthalpic contributions. In this work, we deliberately tuned only the second contribution. We theoretically unveil that $C_{60}$ exists as a partially charged particle in many solutions. This phenomenon can be employed to engineer more efficient solvents for fullerenes, since obtaining

*a real solution of this compound is of significant interest for many fundamental and applied areas.*

5. Conclusions

To recapitulate, an alternative approach to increase dispersion of $C_{60}$ molecules was outlined in the present work. Although a zero solubility in [BMIM][BF$_4$] was recorded at 300 K (in perfect agreement with experimental observations[52]), the solubility increases drastically with temperature: S (310 K)=5 g l$^{-1}$, S (320 K)=49 g l$^{-1}$, S (333 K)>66 g l$^{-1}$. The accuracy of the results is justified by carefully parameterized non-bonded interactions among all components in the simulated system based on electronic structure calculations. Although our atomistic-precision MD study cannot compete with all-electron description in accuracy, it is the only computationally feasible approach to derive insights for the appropriate time (over 300 ns) and length (over 500 nm$^3$) scales.

Here, we chose [BMIM][BF$_4$] as an exemplary RTIL due to its wide availability and usability in fundamental research and technology.[36, 38, 52, 54] We suppose that certain other RTILs can exhibit competitive affinity towards fullerene (enthalpic factor). As fullerene exhibits π-stacking with aromatic compounds, one can hypothesize that polyaromatic cations, such as isoquinolinium$^+$, isothiouronium$^+$, triphenylphosphonium$^+$, may be decent solvents. In the latter case, both electronic polarization and π-stacking are involved in the intermolecular binding. Unfortunately, we currently do not possess a trustworthy information about liquid ranges of these novel compounds.

Trace admixtures of salts containing alkali ions, such as Li$^+$, Na$^+$, K$^+$, may improve solubility, provided that small, mobile cations localize near the $C_{60}$ surface or penetrate inside the cavity. Previous study suggested that even a single lithium atom inside $C_{60}$ greatly perturbs electronic energy levels of the whole structure.[10] Experimental studies on well-dispersed $C_{60}$ molecules in [BMIM][BF$_4$] and other mentioned RTILs upon a wide temperature range are of large fundamental interest.

**Acknowledgements**

The computations have been partially supported by the Danish Center for Scientific Computing (Horseshoe 5). C. M. and E. E. F. thank Brazilian agencies FAPESP and CNPq for support.